\documentclass[aps,prx,twocolumn,superscriptaddress,showpacs,floatfix]{revtex4-1}
\usepackage{amsmath,amssymb,graphicx}
\usepackage{wasysym}
\usepackage[utf8]{inputenc}
\usepackage[T1]{fontenc}
\usepackage{xcolor}
\usepackage{epstopdf}

\IfFileExists{newtxtext.sty}
   {\usepackage{newtxtext,newtxmath}}
   {\IfFileExists{stix.sty}
      {\usepackage{stix}}
      {\IfFileExists{mathptmx.sty}
      {\usepackage{mathptmx}}{} } }

\usepackage{textcomp}

\usepackage{bm}

\IfFileExists{siunitx.sty}{\usepackage{booktabs,siunitx}}{}

\pdfoutput=1
\usepackage{color}
\definecolor{LinkColor}{rgb}{0.256,0.439,0.588}
\usepackage{hyperref}

\usepackage{pifont}

\begin{document}

\title{Symmetry Enforced Self-Learning Monte Carlo Method Applied to the Holstein Model}

\author{Chuang Chen}
\affiliation{Beijing National Laboratory for Condensed Matter Physics and Institute of Physics, Chinese Academy of Sciences, Beijing 100190, China}
\affiliation{School of Physical Sciences, University of Chinese Academy of Sciences, Beijing 100190, China}
\author{Xiao Yan Xu}
\email{wanderxu@gmail.com}
\affiliation{Department of Physics, Hong Kong University of Science and Technology, Clear Water Bay, Hong Kong, China}
\author{Junwei Liu}
\affiliation{Department of Physics, Hong Kong University of Science and Technology, Clear Water Bay, Hong Kong, China}
\author{George Batrouni}
\affiliation{Universit\'e C\^ote d'Azur, INPHYNI, CNRS, 0600 Nice, France}
\affiliation{Beijing Computational Science Research Center, Beijing, 100193, China}
\affiliation{MajuLab, CNRS-UNS-NUS-NTU International Joint Research Unit UMI 3654, Singapore}
\affiliation{Centre for Quantum Technologies, National University of Singapore, 2 Science Drive 3, 117542 Singapore}
\author{Richard Scalettar}
\affiliation{Physics Department, University of California, Davis 95616 USA}
\author{Zi Yang Meng}
\email{zymeng@iphy.ac.cn}
\affiliation{Beijing National Laboratory for Condensed Matter Physics and Institute of Physics, Chinese Academy of Sciences, Beijing 100190, China}
\affiliation{CAS Center of Excellence in Topological Quantum Computation and School of Physical Sciences,
University of Chinese Academy of Sciences, Beijing 100190, China}

\begin{abstract}
Self-learning Monte Carlo method (SLMC), using a trained effective model
to guide  Monte Carlo sampling processes, is a powerful general-purpose
numerical method recently introduced to speed up simulations in
(quantum) many-body systems. In this work, we further improve the
efficiency of SLMC by enforcing physical symmetries on the
effective model. We demonstrate its effectiveness in the Holstein
Hamiltonian, one of the most fundamental many-body descriptions of
electron-phonon coupling. Simulations of the Holstein model are
notoriously difficult due to the combination of the typical cubic scaling
 of fermionic Monte Carlo and the presence of 
 extremely long autocorrelation times.
Our method addresses both bottlenecks. This enables simulations 
on large lattices in the most difficult parameter
regions, and evaluation of the critical point for the
charge density wave transition at half-filling with high precision.  
We argue that our work opens a new research area of quantum Monte Carlo (QMC), providing a general procedure to deal with ergodicity
in situations involving Hamiltonians with multiple, distinct low energy
states.
\end{abstract}

\date{\today}
\maketitle

{\it Introduction}\,---\, Electron-phonon coupling is ubiquitously
present in condensed matter materials, responsible not only for the nature
of basic, single-particle, properties such as the resistance and 
renormalized quasiparticle mass\cite{giustino2017},
but also for more exotic collective phenomena such as metal-insulator
transitions\cite{han2000}, 
charge density wave (CDW) phases\cite{gruner1985,gruner1988}, and 
superconductivity (SC)\cite{bennemann2001,BCS1957}.  Electron-phonon
coupling also has a rich interplay with electron-electron 
interactions\cite{kulic1994,gebhard1997}.
The Holstein Hamiltonian~\cite{HOLSTEIN1959}, which describes spinful
electrons hopping on a latttice and interacting locally with
a phonon degree of freedom, is one of the most simple models
of this rich physics, incorporating both polaron formation
\cite{kornilovitch98,hohenadler04,ku02,macridin04,bonca99}
in the dilute limit, and collective insulating CDW and SC transitions.
Despite its simplicity,
investigation of the Holstein Hamiltonian is extremely challenging,
especially within an exact treatment of both its
bosonic (phonon) and electronic degrees of freedom.

With the help of the Lang-Firsov transformation,  
QMC studies of Holstein (and related) models in 1D can be performed
for large systems and interesting parameter
regimes~\cite{Hohenadler2007}. The case of retarded interaction in 1D is addressed with directed-loop QMC~\cite{Weber2017SSE}.  
There have also been many attempts to explore the phase diagram of
the 2D Holstein and Holstein-Hubbard
models~\cite{Scalettar1989,Marsiglio1990,Levine1990,Noack1991,Levine1991,Vekic1992,Vekic1993,Niyaz1993,Berger1995,Zheng1997,Nowadnick2012,Nowadnick2015,Johnston2013,Murakami2013,Ohgoe2017,Weber2017,Esterlis2017,Costa2018}.
Recent studies of the metal to CDW phase transition in the weak-coupling
regime~\cite{Weber2017} and the competition between CDW and
superconductivity at intermediate coupling strength with phonon
dispersion~\cite{Costa2018} and different Fermi surface
topologies~\cite{Esterlis2017} have broadened the
understanding of the model and its relevance to the microscopic mechanism of
superconductivity. 

However, for 2D and 3D, these QMC methods are limited
by the necessity of the expensive evaluation of a
fermion determinant which enters the weight of the configuration, and
also long autocorrelation times even away from critical
points~\cite{Hohenadler2008}. Thus, even though a sign 
problem\cite{loh1990,troyer2005}
is absent for the Holstein model, reliable results for the complete phase
diagram, in particular for the most interesting intermediate-coupling
strength of the model are still missing.
The cost of treating fermion determinants is
largely unavoidable.  However, the ergodicity problem has been
successfully solved, on an instance-by-instance basis, in a number of 
classical and QMC approachs~\cite{SwendsenWang1987,Wolff1989,Syljuaasen2002}. It remains the key bottleneck of
many others\cite{ceperley03} 
(apart from the sign problem).  Indeed, the
failure of ergodicity challenges determinant and constrained path
QMC~\cite{assaad08,ShiweiZhang1999}, 
lattice gauge theory simulations\cite{burgio07,amato15}, impurity
solvers\cite{semon14,seth16} in dynamical mean field theory, and configuration interaction
methods in quantum chemistry\cite{shepherd12,thomas14}.

A recently developed self-learning Monte Carlo (SLMC)
method~\cite{liu2016self,liu2016fermion,Xu2017SLMC,Nagaiself2017,Shen2018},
based on a trained effective model to guide the Monte Carlo
simulation~\cite{LiHuang2017a,LiHuang2017b}, shows substantial
improvements over traditional Monte Carlo methods. The central idea of SLMC is to make use of learning algorithms to construct an approximate effective action which can be very rapidly calculated. An exact
simulation is recovered by the evaluation of the full determinant which,
however, can be done relatively infrequently, owing to the accuracy of
the learned effective action. In 2D problems in which fermions
coupled to bosonic fluctuations
exhibit itinerant quantum critical points, $L\times L$ spatial
lattices with $L$ up to 100 can be investigated
at high temperature~\cite{Xu2017SLMC} and $L$ up to 48
has been achieved at low temperatures with $\beta \sim L$ scaling\cite{ZHLiu2017,ZHLiu2018}.

In this Letter, we show how SLMC can be applied to the Holstein Hamiltonian.
Our key results are the following:
(1) long autocorrelation times can be greatly reduced
by designing an effective bosonic Hamiltonian for the
phonon fields which incorporates a global $Z_2$ symmetry
in the original model;
(2) computational complexity is reduced from roughly $O(L^{11})$ to
$O(L^{7})$, i.e., a speedup of $O(L^{4})$ in SLMC over traditional MC
method; 
(3) with such improvements, simulations of lattice sizes up to $L=20$
are possible, allowing
the evaluation of the metal to CDW transition temperature
to an order of magnitude higher accuracy than previously available.
These advantages open a new research area of QMC for strongly correlated systems where SLMC provides powerful and general procedure to improve ergodicity.


{\it Model}\,---\,We study the Holstein Hamiltonian,
\begin{equation}
H = H_{\text{el}} + H_{\text{lat}} + H_{\text{int}},
\label{eq:holsteinHam}
\end{equation}
with
\begin{align}
H_{\text{el}} &=-t\sum_{\langle ij \rangle \sigma}
c_{i \sigma}^{\dagger}c_{j \sigma}^{\phantom{\dagger}}
-\mu\sum_{i \sigma} n_{i \sigma}, \nonumber\\
H_{\text{lat}} &=\sum_{i}\left(\frac{M\Omega^{2}}{2} 
X_{i}^{2}+\frac{1}{2M} P_{i}^{2} \right), \nonumber\\
H_{\text{int}}&=g\sum_{i \sigma} n_{i \sigma} X_{i}.
\end{align}
$H_{\text{el}}$ describes spinful electrons hopping on a 2D square
lattice of linear size $L$, $H_{\text{lat}}$ is the free phonon Hamiltonian,
and $H_{\text{int}}$ describes a local coupling between electron
density and the phonon displacement at site $i$. 
$\Omega$ is the phonon frequency, and $g$ is the electron-phonon
coupling.
We set $M=t=1$ as the units of mass and
energy, and focus on half-filling ($\mu=\frac {g^2} {\Omega^2}$,
$\langle n_i \rangle=1$).  
On a square lattice, 
$W=8t$ is the non-interacting bandwidth and
$\lambda=\frac{g^{2}}{M\Omega^{2}W}=\frac{g^{2}}{8t\,\Omega^{2}}$ 
provides a dimensionless measure of
the electron-phonon coupling. 
In this work, we focus on the
intermediate coupling strength $\lambda=0.5$ ($g=1$, $\Omega=0.5$), 
a parameter regime which is more challenging than that explored
in several recent works~\cite{Weber2017,Costa2018}. 

One can see some of the fundamental physics of the Holstein model by
considering the atomic limit ($t=0$).  Completing the square in
$\frac{\Omega^{2}}{2}X_{i}^{2}+gn_{i}X_{i}$, and integrating out the
phonon degrees of freedom, leads to an effective attraction
$\frac{g^2}{\Omega^2}$ between spin up and spin down electrons, and to
pair formation~\cite{Johnston2013}.  At low temperatures, when $t$ is made nonzero, these
pairs can either organize into an insulating CDW pattern (which tends to
happen at half-filling), or condense into a SC phase (at incommensurate
density).  On the other hand, integrating out the fermions leads to a
potential energy surface for the phonons with two minima, at $X_i=0$ and
$X_i=-\frac{2g}{\Omega^{2}}$, corresponding to empty ($n_i=0$) and double
($n_i=2$) occupation respectively.  The large barrier at single
occupation $n_i=1$ between these minima is the fundamental causes
of long autocorrelation times in QMC simulations.

Several QMC methods have been used to simulate the Holstein
model~\cite{Scalettar1989,Spencer2005,Weber2017,kornilovitch98,hohenadler04,macridin04,berciu10,mckenzie96,Vekic1992,noack93,Marsiglio1990}.
Here we use 
determinant quantum Monte Carlo (DQMC)
~\cite{Blankenbecler1981,Hirsch1985,AssaadEvertz2008}, which is
especially effective in dimension $D>1$ and in the large coupling regime. 
In this approach, the inverse temperature $\beta = L_\tau \Delta\tau$ 
is discretized (we use $\Delta\tau=0.1$ in this work),
and a path integral expression for the partition function
is constructed in terms of the quantum coordinates in space $i \in L^2$ 
and imaginary time index $l=1,2,\cdots L_\tau$.  
The fermions are integrated out, resulting
in a weight $\omega[\mathcal{X}]$ for the phonon fields $X_{i,l}$ 
which consists of a 
product of a bosonic piece $e^{-S_{\rm Bose}\Delta \tau}$ with $S_{\rm Bose}= \frac{\Omega^2}{2} \sum_{i,l} X_{i,l}^2 + 
\sum_{i,l} (\frac{X_{i,l+1}-X_{i,l}}{\Delta\tau})^{2}$ 
and a fermion contribution $\big({\rm det} M(\{X_{i,l}\})\,\big)^2$.
The square comes from the fact that the two spin species couple
to the phonon field in the same way, giving rise to identical 
determinants.
Here $M$ is a matrix of dimension $N=L^2$.

\begin{figure}[htp!]
\includegraphics[width=\columnwidth]{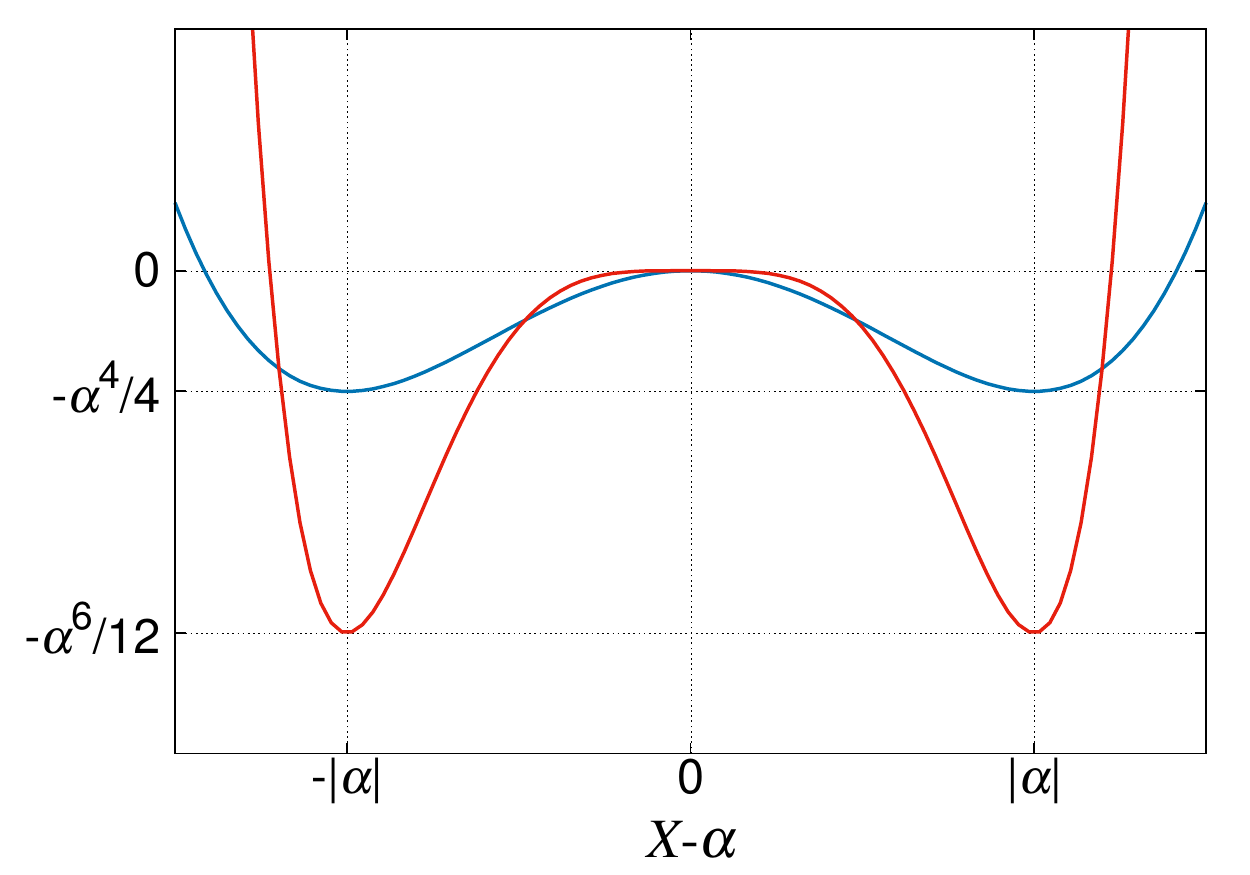}
\caption{The symmetric functions used to construct the phonon potential have
minima at $\pm |\alpha|$ with $\alpha=-\frac{g}{\Omega^2}$. The blue line is
$\frac{1}{4}(X-\alpha)^{4} - \frac {{\alpha}^2} {2} (X - \alpha )^2 $
and red line is  $\frac{1}{6}(X-\alpha)^{6} - \frac {{\alpha}^2} {4} (X
- \alpha )^4 $. }
\label{fig:potential}
\end{figure}

Local updates of a single phonon coordinate $X_{i,l}$ can be done
with a computation cost $O(N^2)$, so that
a sweep through all $NL_\tau$ components scales as 
$N^3 L_\tau = L^6 L_\tau$ in $D=2$.
The first computational bottleneck, associated with the fermionic
degrees of freedom is immediately evident:  Doubling the
linear lattice size $L$ results in a 64-fold increase in
computation time in $D=2$. 
Part of the origin of the second bottleneck is also clear from
the form of $S_{\rm Bose}$.  The phonon degrees of freedom 
on adjacent imaginary time slices $l$ are tightly coupled,
especially so as $\Delta \tau$ becomes small.  Moves of a single
coordinate are thus energetically unfavorable.
A block update of all the imaginary time
phonon coordinates $l$ of a single spatial lattice site $i$
helps surmount this problem.   It is also important to
tune the value of the change in the block update 
$\Delta X =\frac{2g}{\Omega^2}$, to shift the phonon fields between
the two minima. More details can be found in Sec.~\ref{sec:DQMC} of the
Supplemental Materials (SM)~\cite{suppl}.

\begin{figure}[tp!]
\includegraphics[width=\columnwidth]{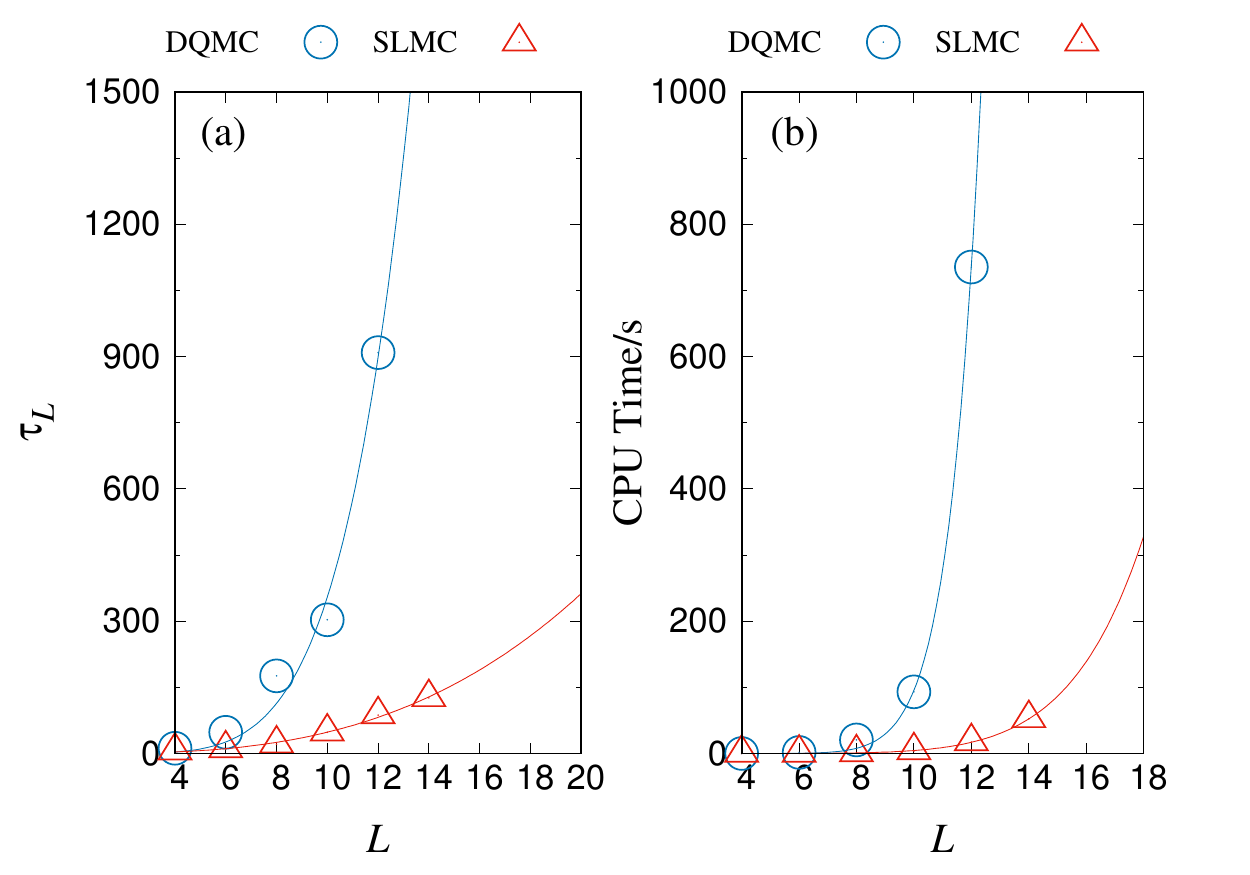}
\caption{(a) Comparison of autocorrelation time of the CDW structure
  factor versus $L$ for DQMC and SLMC. Simulations were done
 at the critical point $T_c$ for the CDW transition. SLMC
  greatly suppresses the autocorrelation time, with dynamical exponent
  $z\sim 2.9$, while for DQMC, $z\sim 5.1$. (b) Comparison of CPU time
  to obtain one statistically independent configuration 
($\tau_L$ sweeps)
  between DQMC and SLMC.
  Power-law fitting gives $\sim L^{11}$ for DQMC and $\sim
  L^{7}$ for SLMC.  Including the prefactor, SLMC provides
a $\times 50$ speedup for $L=12$ and $\times 300$ speedup for
  $L=20$.}
\label{fig:autocorr}
\end{figure}

However, even with the local and block updates, the autocorrelation time
$\tau_L$
of DQMC for the Holstein model is still found to increase rapidly with
system size, as shown in Fig~\ref{fig:autocorr} (a). 
(Further aspects of Fig~\ref{fig:autocorr} will be discussed later.)
We find $\tau_{L} \sim L^{5.1}$, much worse than the
dynamic critical exponent $\tau_L \sim L^z$ with $z=2$ associated
with classical Monte Carlo simulations with local update, e.g.~of the Ising
model near its critical point\cite{SwendsenWang1987}.  
Such autocorrelation times lead to a
situation where in 2D,   
$L \sim 10-14$ is at the limit of DQMC simulations.
It is important to note that large $\tau_L$ occurs even away
from any critical point, but in this work we focus on the most difficult situation -- critical slowing down near $T_c$.

{\it SLMC}\,---\,To overcome these problems, we apply SLMC 
to the Holstein model. 
The first step\cite{liu2016self,liu2016fermion,Xu2017SLMC,Nagaiself2017,Shen2018},
is to obtain an effective model by self-learning on configurations
generated with DQMC updates according to Eq.~\ref{eq:holsteinHam}. Here, at each temperature studied, we use $80,000$ configurations obtained for $L=6$ systems to train the effective model $H^{\text{eff}}$, which we choose to have polynomial form,
\begin{equation}
H^{\text{eff}}[X] =  E_0 +  J_i X_i + J_{ij}X_i X_j + \cdots
\label{eq:effectiveHam}
\end{equation}
where $E_0$ is the zeroth order background, $J_i$ are the first order
terms, $J_{ij}$ are second order terms, $\cdots$, and indices $i$ and
$j$ are now combined
space-imaginary time coordinates. Such a form is very natural, as after
tracing out the fermions, the bosonic fields acquire long-range
interactions in space-time beyond the bare level. One can make use of the symmetry of the original model to further
reduce the number of parameters in the effective model, consequently reduce the effort and uncertainty in the fitting step.

The two potential minima of the Holstein model are symmetric with 
respect to $X=-\frac g {\Omega^2}\equiv \alpha$. 
We build this 
into the effective model by representing the potential by
functions with these two minima (Fig.~\ref{fig:potential} and see Sec.~\ref{sec:DEFF} of the SM for details). 
We find that for the phonon fields in the Holstein model, two functions
are sufficient to fit an appropriate barrier 
width and height. Our effective model thus has the
following form (up to a constant),
\begin{align}
-\beta H^{\text{eff}} & = 
J_{k}\sum_{i\tau}(X_{i\tau+1}-X_{i\tau})^{2} 
\nonumber \\ 
&+ J_{p}\sum_{i\tau}\left(\frac{1}{4}(X_{i\tau}-\alpha)^{4} - \frac
{{\alpha}^2} {2} (X_{i\tau} - \alpha )^2 \right)    
\nonumber  \\ 
&+ J_{p}'\sum_{i\tau}\left(\frac{1}{6}(X_{i\tau}-\alpha)^{6} - \frac
{{\alpha}^2} {4} (X_{i\tau} - \alpha )^4 \right)   
\nonumber  \\ 
 &+ J_{nn}\sum_{\langle ij \rangle
  \tau}(X_{i\tau}-\alpha)(X_{j\tau}-\alpha) 
\nonumber \\ 
&+  J_{nn}'\sum_{i\langle \tau\tau' \rangle}(X_{i\tau}-\alpha
)(X_{i\tau'}-\alpha), 
\label{eq:effectiveHamHolsteinmain}
\end{align}
where the $J_k$-term comes from the phonon kinetic energy,  $J_p$ and
$J_p'$-terms are functions which produce the two global minima of
Fig.~\ref{fig:potential} and $J_{nn}$ and $J'_{nn}$ are the
nearest neighbor interaction in the spatial and temporal directions,
respectively. Longer range interactions are found to contribute little
to the weight and are thus omitted (we have tried spatial interaction to $L/2$ and temporal interaction encompassing 10 time slices). One key remark here is this
effective model captures a global $Z_2$ symmetry, namely, a
global mirror operation on $X$ with axis $\alpha$ leaving
$H^{\text{eff}}$ invariant.

With the effective model in the form of
Eq.~\ref{eq:effectiveHamHolsteinmain}, the training procedure is
straightforward. Given a configuration $\mathcal{X}$ of the phonon
field and its corresponding weight $\omega[\mathcal{X}]$, generated in
DQMC, we have
\begin{equation}
-\beta H^{\text{eff}} [\mathcal{X}] = \ln \left(  \omega[\mathcal{X}]
\right). 
\label{eq:fitting}
\end{equation}
Combining Eq.~\eqref{eq:effectiveHamHolsteinmain} and
Eq.~\eqref{eq:fitting}, optimized values of $J_k$, $J_p$, $J_p'$,
$J_{nn}$ and $J_{nn}'$ (shown in Tab.~\ref{table1} in the SM) can be
readily obtained through a multi-linear regression
~\cite{liu2016self,liu2016fermion,Xu2017SLMC} using all the
configurations prepared with DQMC. Note for each temperature, we only train $H^{\text{eff}}$ from small system size ($L=6$), but use it to larger systems (up to $L=20$) in SLMC. 

We use the effective model to guide the Monte
Carlo simulation of the original model, namely, propose 
many updates of the phonon fields according to 
Eq.~\ref{eq:effectiveHamHolsteinmain}, this is the so-called cumulative update in SLMC~\cite{liu2016fermion,Xu2017SLMC}. We then calculate the
acceptance ratio of the final phonon field configuration via the
expensive fermion determinant only rarely. 
There are two advantages of SLMC over
DQMC. First, the effective model is purely bosonic and its local
update is $O(1)$ since it bypasses the calculation of fermion
determinants. Second, since the effective model is bosonic, global
updates, such as Wolff and other cluster update
schemes~\cite{SwendsenWang1987,Wolff1989}, are easy to implement.
This is crucial since cluster updates in conventional DQMC actually worsen 
the scaling from $O(N^3 L_\tau)$ to $O(N^4 L_\tau)$.

The Holstein model exhibits a finite temperature metal to
CDW insulator phase transition at half-filling  belonging to the 2D
Ising universality class~\cite{Weber2017,Costa2018}.
As discussed in detail in Sec.~\ref{sec:Wolff} of the SM, we designed a
modified Wolff cluster update on the effective model, by building the
cluster in space and including all sites of temporal columns which 
addresses additional long autocorrelation times associated with
proximity to the critical point. 
This
modified Wolff update successfully
reduces the autocorrelation time of Monte Carlo simulations from $L^{5.1}$ to $L^{2.9}$ (as shown in
Fig.~\ref{fig:autocorr}(a)). So the dynamical exponent is reduced by
$\Delta z \ge 2$, an equivalent improvement to that provided by
cluster moves in the classical Ising model~\cite{SwendsenWang1987,Wolff1989}.

Using this combination of updates
on the effective model, we propose cumulative move~\cite{liu2016fermion,Xu2017SLMC} of the phonon
field for the original model, combined with a final
acceptance ratio,
\begin{equation}
 A(\mathcal{X} \rightarrow \mathcal{X}') = \min \left\{1, \ \frac
 {\exp\left(-\beta H[\mathcal{X}'] \right) } { \exp\left(-\beta
   H[\mathcal{X}]\right)} \frac {\exp\left(-\beta H^{\text{eff}}
   [\mathcal{X}] \right)} {\exp\left(-\beta H^{\text{eff}}
   [\mathcal{X}'] \right)} \right\},
\label{eq:acceptanceratio}
\end{equation}
which ensures detailed balance and hence simulation of the original Holstein $H[\mathcal{X}]$.

\begin{figure}[htp!]
\includegraphics[width=\columnwidth]{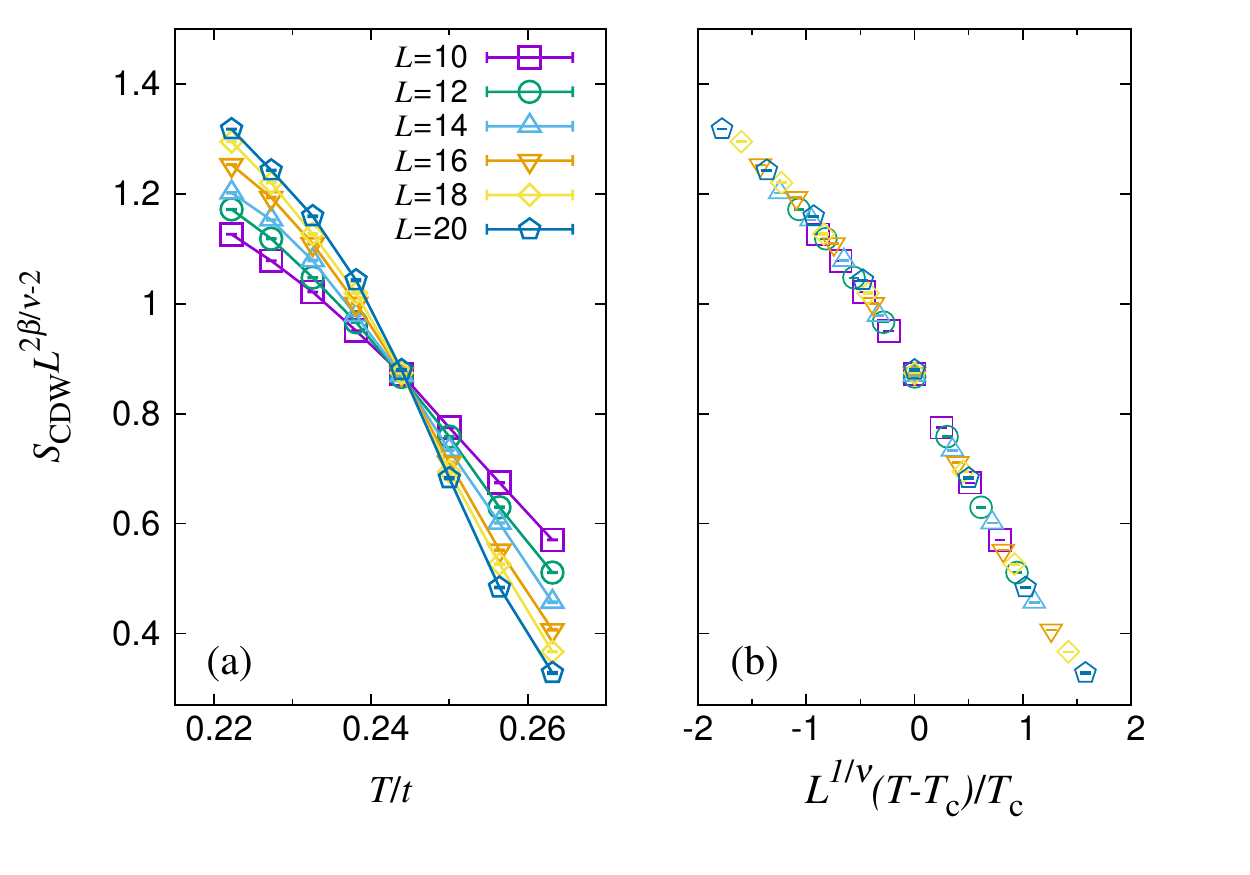}
\caption{(a) Finite size scaling analysis showing $S_{\text{CDW}}
  \ L^{-7/4}$ versus $T/t$. The critical point $T_c=0.244(3)$ is
  determined from the crossing of finite size data. (b) Data collapse
  of $S_{\text{CDW}} \ L^{-7/4}$ versus $L^{1/\nu} \ (T-T_c)/T_c$ with
  $\nu=1$. Note the quality of the collapse is better than that in Ref.~\cite{Costa2018}, due to larger $L$ obtained with SLMC.}
\label{fig:cross}
\end{figure}

{\it Results}\,---\, To compare SLMC and DQMC, Fig.~\ref{fig:autocorr}
(a) depicts the autocorrelation time of the CDW structure factor 
$S_\text{CDW} = \frac 1 {L^2} \sum_{ij} (-1)^{i+j} 
\left( \langle n_i n_j\rangle - 
\langle n_i\rangle \langle n_j \rangle \right)$ 
for DQMC and SLMC.  We have chosen the most challenging criterion,
both by analyzing a long range quantity associated with the order
parameter (which has much longer correlations than simple local
quantities like the energy) and also by tuning the temperature to
$T=T_c$. To compare in an equal footing, a MC step in DQMC is defined as a sweep with local update plus 4 block updates; whereas a MC step in SLMC is defined as local plus Wolff-cluster updates. As can be seen from the fitting of the $\tau_L \sim L^z$, severe critical slowing down is observed in DQMC, with dynamical exponent
($z\sim 5.1$); on the other hand, $z\sim 2.9$ in SLMC. A reduction of $\Delta \ z \ge 2$ is achieved -- equivalent to the improvement of cluster over local moves in classical Ising model~\cite{SwendsenWang1987,Wolff1989}.   

In Fig.~\ref{fig:autocorr}(b), we show
how much  CPU time is needed to obtain a statistically
independent phonon configuration.  SLMC 
provides a $\times 50$ speedup for
$L=12$, and more than $\times 300$ speedup for $L=20$.
With this dramatic improvement, we are able to simulate
the Holstein model in the difficult parameter regime
($\lambda \sim 0.5$) and determine the critical point with high
precision. 
At the critical point, the
finite size scaling behavior $S_\text{CDW}/L^{2} = L^{-2\beta/\nu}
f\left(L^{1/\nu}\left(\frac {T-T_c} {T_c}\right)\right)$ is expected,
where $\beta=\frac{1}{8}$, $\nu=1$ are the 2D Ising critical
exponents. Fig.~\ref{fig:cross}(a) shows $L^{-7/4} S_{\text{CDW}}$
versus $T$ for $L=10$ to $20$. 
Their crossing point yields the critical temperature
$T_c=0.244(3)$.  In Fig.~\ref{fig:cross}(b), we further rescale
the horizontal-axis with $L\left(\frac {T-T_c} {T_c}\right)$ giving an excellent
data collapse. 
Our value of $T_c$ represents a substantial improvement
over existing work which typically reports 
maximal lattice sizes of $L=10 \sim 12$~\cite{Costa2018}.

{\it Conclusions}\,---\,In this Letter, we applied SLMC to simulations
of the electron-phonon interaction in the Holstein model,
where QMC is very difficult due to long autocorrelation times and
the expense of the fermion determinant evaluation.
By imposing a global
$Z_2$ symmetry in the effective model of SLMC, we have successfully
captured the global minima of the phonon potential. In addition, we
designed a Wolff-cluster update in the effective model which
greatly reduces the autocorrelation time challenge which has hampered
simulation of the Holstein model for several decades.

With these improvements on the effective model, SLMC simulation of
the Holstein Hamiltonian can be pushed to larger system sizes,
comparable with other interacting fermion 
systems~\cite{Varney2009,Meng2010,Toldin2015,YQQin2017} or itinerant
quantum criticality models~\cite{Xu2017,ZHLiu2017,ZHLiu2018}). 
This allows for much more reliable determination of critical properties.

The idea of imposing symmetry in the effective model can be
generalized to other situations. For example, in the Hubbard and
Hubbard-like models, one can introduce a continuous auxiliary field
$\phi$ to decouple the interaction in the spin channel, $\alpha \phi
(n_{\uparrow} -n_{\downarrow})$.  
High barriers interfere with movement between minima at
$n_\uparrow-n_\downarrow = \pm 1$. 
Block updates in DQMC can be used \cite{Scalettar1991}
but are very time consuming, especially as the lattice
size increases.   Our work suggests that SLMC can improve this situation by
proposing global updates based on effective models 
like Eq.~\eqref{eq:effectiveHamHolsteinmain} with 
appropriate symmetry to
capture the minima and lead to a big reduction in
autocorrelation time.  
Construction of large scale moves is not only computationally
much inexpensive within the context of effective models
like Eq.~\eqref{eq:effectiveHamHolsteinmain}, but it is also much easier to 
incorporate intuitive physical pictures of the low energy configurations.
This suggests SLMC will provide a general framework for
improving ergodicity beyond the Holstein model illustrated here.

If successful, outstanding 
problems such as spectral properties in the Mott insulator, or the recently
discovered phenomena of symmetric mass
generation~\cite{He2016,YiZhuangYou2017} could be explored more
efficiently.  Potential applications extend 
outside condensed matter physics to, for example, QMC simulations
of shell model Monte
Carlo~\cite{Koonin1997} in high-energy physics. 
The key requirement for our application of SLMC is only that the
interactions can be decoupled to continuous bosonic fields associated
with fermion bilinears, where several minima are present. 



{\it Acknowledgments}\,---\, C.C. and Z.Y.M. acknowledge the valuable discussions with Martin Hohenadler and Fakher Assaad on the Holstein model, and thank the support
from the Ministry of Science and Technology of China under Grant
No. 2016YFA0300502, the key research program of the Chinese Academy of
Sciences under Grant No. XDPB0803, the National Natural Science
Foundation of China under Grants No. 11421092, 11574359 and 11674370,
and the National Thousand-Young Talents Program of
China. X.Y.X. acknowledges the support of HKRGC through grant
C6026-16W and also gratefully acknowledges the hospitality of Institute of
Physics, Chinese Academy of Sciences. We thank the following
institutions for allocation of CPU time: the Center for Quantum
Simulation Sciences in the Institute of Physics, Chinese Academy of
Sciences; the Tianhe-1 platform in the National Supercomputer Center
in Tianjin. GGB acknowledges support from the Universit\'e C\^ote d'Azur IDEX Jedi.
The work of RTS is supported by the U.S.~Department of Energy under
grant DE-SC0014671.

\bibliography{main}

\clearpage

\appendix
\begin{center}
\textbf{\large Supplemental Material: Symmetry Enforced Self-Learning Monte Carlo Method Applied to the Holstein Model}
\end{center}
\setcounter{equation}{0}
\setcounter{figure}{0}
\setcounter{table}{0}
\setcounter{page}{1}
\makeatletter
\renewcommand{\thetable}{S\arabic{table}}
\renewcommand{\theequation}{S\arabic{equation}}
\renewcommand{\thefigure}{S\arabic{figure}}

\section{DQMC for the Holstein model}
\label{sec:DQMC}


As discussed in the main text, the Holstein model describes
electrons hopping on a lattice and interacting with local
phonon modes. Before introducing the implementation of SLMC
on the Holstein model, we first give a short review of the DQMC
algorithm, based mainly on Ref.~\cite{Johnston2013}.

We define $K\equiv H_{\text{el}}+H_{\text{lat}}$ as the
non-interacting terms for the electron and lattice (phonon) degrees of
freedom.  To implement DQMC, we start with partition function
\begin{equation}
Z=\text{Tr}(e^{-\beta H})=\text{Tr}(e^{-\Delta\tau
  H_{\text{int}}}e^{-\Delta\tau K})^{L} + O((\Delta\tau)^2).
\end{equation}
The Trotter-Suzuki decomposition is performed by discretizing $\beta$ 
into $L_\tau$ segments with $\Delta \tau=\frac{\beta}{L_\tau}$.

By tracing out the fermions, the partition function is expressed as an
integral over the phonon fields,
\begin{equation}
Z=\int dX
e^{-S_{\text{Bose}}\Delta\tau} \, \text{det}M_{\uparrow}
\, \text{det}M_{\downarrow},
\end{equation}
where $M_{\sigma} = I + B_L^\sigma B_{L-1}^\sigma \cdots B_1^\sigma$
with $B_{l}^{\uparrow(\downarrow)}=e^{-\Delta\tau gX(l)}e^{-\Delta\tau
  K}$,
$S_{\text{Bose}}=\frac{\Omega^{2}}{2}\sum_{i,l}X_{i,l}^{2}+\sum_{i,l}(\frac{X_{i,l+1}-X_{i,l}}{\Delta\tau})^{2}$.

The key quantity in DQMC is the single-particle Green function 
\begin{equation}
[G^{\sigma}(l)]_{ij}=[I+B_{l}^{\sigma}...B_{1}^{\sigma}B_{L}^{\sigma}...B_{l+1}^{\sigma}]_{ij}^{-1},
\end{equation}
which
is used to evaluate the acceptance ratio and to obtain physical
observables. 
The ratio of fermion determinants 
\begin{equation}
R=R^{\uparrow}R^{\downarrow}=
\frac{\text{det}M'^{\uparrow}\text{det}M'^{\downarrow}}
     {\text{det}M^{\uparrow}\text{det}M^{\downarrow}},
\end{equation}
is used to accept or reject updates to the phonon field.
If the update is local, a fast $O(1)$ evaluation of $R$ is possible:
\begin{equation}
R^{\sigma}=1+(1-[G^{\sigma}(l)]_{ii}) [\Delta^{\sigma}(i,l)]_{ii}.
\end{equation}
However, the change in the phonon field alters the Green function.
For local updates an $O(N^2)$ procedure is provided by the
Sherman-Morrison formula (compared to an $O(N^3)$ scaling
of a direct recalculation of $G$).
Since only one $B$ matrix is changed from $B^{\sigma}(l)$ to
$B^{\sigma'}(l)=[I+\Delta^{\sigma}(i,l)]B^{\sigma}(l)$, where
$\Delta^\sigma(i,l)$ only has one non-zero element,
$[\Delta^{\sigma}(i,l)]_{jk}=\delta_{ik}\delta_{jk}[\exp(-g\Delta\tau\Delta
  X_{i,l})-1]$,  and
\begin{equation}
[G^{\sigma}(l)]' = G^{\sigma}(l)-
\frac{G^{\sigma}(l)\Delta^{\sigma}(i,l) [I-G^{\sigma}(l)]}{1+
  [1-G_{ii}^{\sigma}(l)]\Delta_{ii}^{\sigma}(i,l)}.
\end{equation}

As discussed in the main text, to overcome the barrier of two minima
in the phonon potential, a block update is applied. By changing the
phonon coordinate for all time slices of one site uniformly through a
reflection with respect to the average phonon displacement
($X(i,\tau)\rightarrow-X(i,\tau)-2g/\Omega^{2}$).

\section{SLMC for the Holstein model}
\label{sec:SLMC}
SLMC~\cite{liu2016self,liu2016fermion,Xu2017SLMC,Shen2018}, based on a
trained effective model to guide Monte Carlo simulation, is proposed
as a general method to simulate (quantum) many-body systems.  As
described in Ref~\cite{liu2016self,liu2016fermion,Xu2017SLMC} and in
the main text, SLMC is comprised of four steps. In the case of Holstein model discussed in the work. We first generate $80,000$
configurations of small size $L=6$ from DQMC. Second, at each temperature ($\beta$ from 3.8 to 4.5, with $\Delta\tau=0.1$ and only considering 10 time slices in the training) we train an
effective model by using data from the
first step. Third, we simulate the effective model with many local and
global moves, i.e., cumulative updates. Finally, the proposed updates of the effective model
are accepted/rejected by applying detailed balance, described in
Eq.~\ref{eq:acceptanceratio}, of the original model. In general, as
shown here, as long as the effective model is a good description of
the original Hamiltonian, the speedup of SLMC over conventional MC
methods can be substantial. This leads us to the next section on how
to design a good effective model.

\section{Designing the effective model}
\label{sec:DEFF}
A good effective model is essential for SLMC. As described in Appendix
A, after tracing out the fermion degrees of freedom in
DQMC, one obtains an expression for the partition function
involving only an integration over the phonon fields degree of freedom. 
However, to evaluate the determinant which enters the resulting weight
for the phonons is numerically very slow.  This is the reason we
need a simpler, bosonic effective model to accelerate this step.
We consider the atomic limit, where the phonon fields can be isolated
on single sites, then the phonon field potential has the form
$\frac{\Omega^{2}}{2}X_{i}^{2}+gn_{i}X_{i}-\mu n_{i}$. As discussed in
the main text, when $g\neq0$, it is easy to see there are two
potential minima at $X_i=0$ and $X_i=-\frac{2g}{\Omega^{2}}$, with
electron filling $n_i=0$ and $n_i=2$, respectively. Note that there is
also a maximum in the middle $X_i=-\frac g {\Omega^2}\equiv \alpha$. 
Integrating the function $X_i(X_i-2\alpha)(X_i-\alpha)^a$ with
odd $a$ will give exactly the shape of two minima and one maximum in
the phonon potential. This is how we obtained the functional forms
$\frac{1}{4}(X_{i\tau}-\alpha)^{4} - \frac {{\alpha}^2} {2} (X_{i\tau}
- \alpha )^2$ and $\frac{1}{6}(X_{i\tau}-\alpha)^{6} - \frac
{{\alpha}^2} {4} (X_{i\tau} - \alpha )^4$ in the $J_p$ and $J'_p$
terms in the effective model in Eq.~\ref{eq:effectiveHamHolsteinmain}
in the main text.

By also considering the momentum term for phonons, the spatial
and temporal interaction terms among the phonons, the effective
Hamiltonian with $a=1,3$ terms takes the form
\begin{eqnarray}
-\beta H^{\text{eff}} & = &
J_{k}\sum_{i\tau}(X_{i\tau+1}-X_{i\tau})^{2} \nonumber \\ &&
+ J_{p}\sum_{i\tau}\left(\frac{1}{4}(X_{i\tau}-\alpha)^{4} - \frac
{{\alpha}^2} {2} (X_{i\tau} - \alpha )^2 \right)  \nonumber \\ &&
+ J_{p}'\sum_{i\tau}\left(\frac{1}{6}(X_{i\tau}-\alpha)^{6} - \frac
{{\alpha}^2} {4} (X_{i\tau} - \alpha )^4 \right)  \nonumber \\ & &
+ J_{nn}\sum_{\langle ij \rangle
  \tau}(X_{i\tau}-\alpha)(X_{j\tau}-\alpha) \nonumber \\ &&
+ J_{nn}'\sum_{i\langle \tau\tau' \rangle}(X_{i\tau}-\alpha
)(X_{i\tau'}-\alpha).
\label{eq:effectiveHamHolstein}
\end{eqnarray}

With Eq.~\ref{eq:effectiveHamHolstein}
for the effective model, and configurations
generated from DQMC, we perform multi-linear regression and obtain
values of $J_k$, $J_p$, $J_p'$, $J_{nn}$ and $J_{nn}'$. As an example,
Table~\ref{table1} lists the values for $L=6$, $\beta=\beta_c=4.1$,
$\lambda =0.5$.
\begin{table}[ht]
    \caption{Optimized values of  $J_k$, $J_p$, $J_p'$, $J_{nn}$ and
      $J_{nn}'$  obtained through a multi-linear regression
      ~\cite{liu2016self,liu2016fermion,Xu2017SLMC} using all the
      configurations prepared with DQMC on parameters $L=6$,
      $\beta=4.1$, $\lambda=0.5$. } 
    \centering 
    \begin{tabular}{c c c c c} 
    \hline\hline 
    $J_k$ & $J_p$ & $J_p'$ & $J_{nn}$ & $J_{nn}'$ \\ [0.5ex] 
    \hline 
     5.00E1  &  1.39E-2  &  -3.05E-4  &  7.17E-3  &  7.67E-2  \\ [1ex] 
    \hline 
    \end{tabular}
    \label{table1} 
    \end{table}
    
The effective model at other temperatures, is obtained in the similar manner.


\section{Wolff update of the effective Hamiltonian}
\label{sec:Wolff}
In this last section, we discuss how to build the modified
Wolff-cluster update of the effective model. In the Wolff-cluster
update for the 2D or 3D Ising model, the probability of adding a spin
to the cluster is $P_{\text{add}}=1-e^{-2|J|\beta}$, where $2|J|$ is the
energy cost when breaking a bond. For the effective model of the
Holstein Hamiltonian, the phonon fields are continuous. The Wolff
update for the effective model is therefore similar to that in the XY
model~\cite{HASENBUSCH1990}, where the probability of adding a field to the cluster is
\begin{equation}
P_{\text{add}}=1- \exp[-2\beta(\hat{n}\cdot s_{i}) (\hat{n}\cdot
  s_{j})].
\end{equation}
In the Holstein model, for the phonon fields, the probability of adding a field to the cluster is 
\begin{equation}
P_{\text{add}}=1-\exp\left(2\Delta\tau\sum_{\tau}J_{nn}(X_{i\tau}-\alpha)
(X_{j\tau}-\alpha)\right).
\label{eq:Wolffprob}
\end{equation}
Since there are very strong interactions along the temporal direction
($J_k$ is more than $10^3$ times larger than $J_{nn}$), we will only
use the $J_{nn}$ term to build clusters in spatial planes and include
sites in the entire temporal column in the cluster, hence the sum
over $\tau$ in Eq.~\ref{eq:Wolffprob}. We then reflect all phonon
fields in the cluster with respect to the symmetry axis $X=\alpha$. As
the effective model has a global $Z_2$ symmetry about the same
axis, the acceptance ratio of the cluster update is one. In addition
to the Wolff-cluster update, we also sweep over the space-time lattice
of the phonon fields with local updates.  The combination defines the
global move proposal of the phonon field for the original model.
\end{document}